\documentclass[aps,prl,onecolumn,preprint,superscriptaddress,amsmath,amssymb,showpacs]{revtex4-1}
\usepackage{tipa}
\usepackage{amssymb}

\usepackage{graphicx,color}
\usepackage{dcolumn}
\usepackage{bm}

\begin{document}

\title{Determination of Carrier-Envelope Phase of Relativistic Few-Cycle Laser Pulses by Thomson Backscattering Spectroscopy}
\author{M. \surname{Wen}}
\affiliation{State Key Laboratory of Nuclear Physics and Technology,
Peking University, Beijing 100871, China} \affiliation{Institute of
Photonics $\&$ Photon-Technology, Northwest University, Xi'an
710069, China}

\author{L.L. \surname{Jin}}
\affiliation{Department of Physics, Northwest University, Xi'an
710069, China}

\author{H.Y. \surname{Wang}}
\affiliation{State Key Laboratory of Nuclear Physics and Technology,
Peking University, Beijing 100871, China}

\author{Z. \surname{Wang}}
\affiliation{State Key Laboratory of Nuclear Physics and Technology,
Peking University, Beijing 100871, China}

\author{Y.R. \surname{Lu}}
\affiliation{State Key Laboratory of Nuclear Physics and Technology,
Peking University, Beijing 100871, China}

\author{J.E. \surname{Chen}}
\affiliation{State Key Laboratory of Nuclear Physics and Technology,
Peking University, Beijing 100871, China}

\author{X.Q. \surname{Yan}}
\email{x.yan@pku.edu.cn} \affiliation{State Key Laboratory of
Nuclear Physics and Technology, Peking University, Beijing 100871,
China}\affiliation{Key Lab of High Energy Density Physics
Simulation, CAPT, Peking University, Beijing 100871, China}

\date{\today}

\begin{abstract}
A novel method is proposed to determine the carrier-envelope phase
(CEP) of a relativistic few-cycle laser pulse via the central
frequency of the isolated light generated from Thomson
backscattering (TBS). We theoretically investigate the generation of
a uniform flying mirror when a few-cycle drive pulse with
relativistic intensity ($I > 10^{18} {{\rm{W}} \mathord{\left/
 {\vphantom {{\rm{W}} {{\rm{cm}}^{\rm{2}} }}} \right.
 \kern-\nulldelimiterspace} {{\rm{cm}}^{\rm{2}} }}$) interacts with a
 target combined with a thin and a thick foil. The central frequency of the
isolated TBS light generated from the flying mirror shows a
sensitive dependence on the CEP of the drive pulse. The obtained
results are verified by one dimensional particle in cell (1D-PIC)
simulations.

\begin{description}
\item[PACS numbers]
52.38.Ph, 41.75.Jv, 52.59.Ye, 42.30.Rx
\end{description}
\end{abstract}

\maketitle

The developments of laser technology provide the possibility to
create both the ultrashort and the ultraintense laser sources.
Progress in ultrafast laser technology has made it possible to
produce laser pulses with only a few cycles in
duration~\cite{Baltuska}, which gives a way to attosecond physics
and high-order harmonic generation~\cite{Krausz}. Meanwhile,
the availability of superintense laser pulses opens a
window to the physical phenomena occurring in the relativistic and
ultra-relativistic domain~\cite{Mourou}. With the broad bandwidth
material-Ti:sapphire, even few-cycles pulse with peak intensities
exceeding 1 TW/cm$^2$ and duration of 10 fs or shorter are
produced~\cite{Stingl}. The focusing of the few-cycle pulse can
reach ${\rm{ \gg  10}}^{{\rm{18}}} {{\rm{W}} \mathord{\left/
 {\vphantom {{\rm{W}} {{\rm{cm}}^{\rm{2}} }}} \right.
 \kern-\nulldelimiterspace} {{\rm{cm}}^{\rm{2}} }} $
 relativistic intensity on the
 target~\cite{Tsung}, which is suitable for laser wakefield acceleration regime
 to generate monoenergetic electrons~\cite{Schmid} as well as for high harmonic
generation on plasma surfaces~\cite{Heissler} and gas jets. For
relativistic few-cycle laser, carrier-envelop phase (CEP)
measurements are still envisaged for CEP stabilization that will be
necessary to generate single attosecond bursts.

The electric field of a laser pulse can be written as $ E\left( t
\right) = E_0 \left( t \right)\cos \left( {\omega_L \tau + \phi }
\right) $, with $E_0\left( t \right)$ being the pulse envelope,
$\omega_L$ being the frequency of the carrier wave, and $\phi$ being
the CEP~\cite{Dietrich}. The CEP $\phi$ is defined as the offset
between the optical phase and the maximum of the wave envelope of an
optical pulse. The CEP may affect many processes involving
instantaneous laser-matter interaction. On one hand, for few-cycle
pulses, it has been proved that the electric field as a function of
time depends on the CEP, although the envelope is the same for all
pulses. The CEP effects of ultrashort laser pulses are widely
investigated from the non-ionizing optics regime~\cite{Mehendale} to
the ionizing intensity regime~\cite{Paulus}, even to the
relativistic regime~\cite{Nerush}. On the other hand, with a method
for measuring the CEP of a many-cycle pulse \cite{Tzallas}, CEP
effects by intense multi-cycle pulses are experimentally observed
\cite{Jha}.

So far, a method known as stereo above threshold ionization (ATI)
has been demonstrated experimentally to determine the CEP of
few-cycle pulses with intensities up to $I = 10^{14} - 10^{15}
{{\rm{W}} \mathord{\left/
 {\vphantom {{\rm{W}} {{\rm{cm}}^{\rm{2}} }}} \right.
 \kern-\nulldelimiterspace} {{\rm{cm}}^{\rm{2}} }}
$~\cite{Paulus1}, at a precision of about $ {\pi \mathord{\left/
 {\vphantom {\pi  {300}}} \right.
 \kern-\nulldelimiterspace} {300}}
$~\cite{Wittmann}. Other methods of measuring the CEP are possible
through an attosecond photon probe~\cite{Goulielmakis1} and
detection of THz emission generated in a plasma~\cite{Kress}.
However, these methods are not available for laser pulses of
intensities above $I = 10^{16}{{\rm{W}} \mathord{\left/
 {\vphantom {{\rm{W}} {{\rm{cm}}^{\rm{2}} }}} \right.
 \kern-\nulldelimiterspace} {{\rm{cm}}^{\rm{2}} }}$, when relativistic effects become increasingly
 important. Recently, a quantum method is proposed to determine the CEP of
 ultra-relativistic intensity by detecting the angular emission range via
 multiphoton Compton scattering~\cite{Mackenroth}, which is available when the intensity $I >
10^{20}{{\rm{W}} \mathord{\left/
 {\vphantom {{\rm{W}} {{\rm{cm}}^{\rm{2}} }}} \right.\kern-\nulldelimiterspace} {{\rm{cm}}^{\rm{2}}
 }}$.
This Letter reports  the CEP of a relativistic intense ($I
> 10^{18}{{\rm{W}} \mathord{\left/
 {\vphantom {{\rm{W}} {{\rm{cm}}^{\rm{2}} }}} \right.\kern-\nulldelimiterspace} {{\rm{cm}}^{\rm{2}}
 }}$) few-cycle laser pulse can be determined by detecting the spectroscopy of the
isolated Thomson Backscattering (TBS) pulse, which is testified by
an analytical model and particle in cell simulations.

\begin{figure}
\begin{center}
\includegraphics[width=0.8\columnwidth]{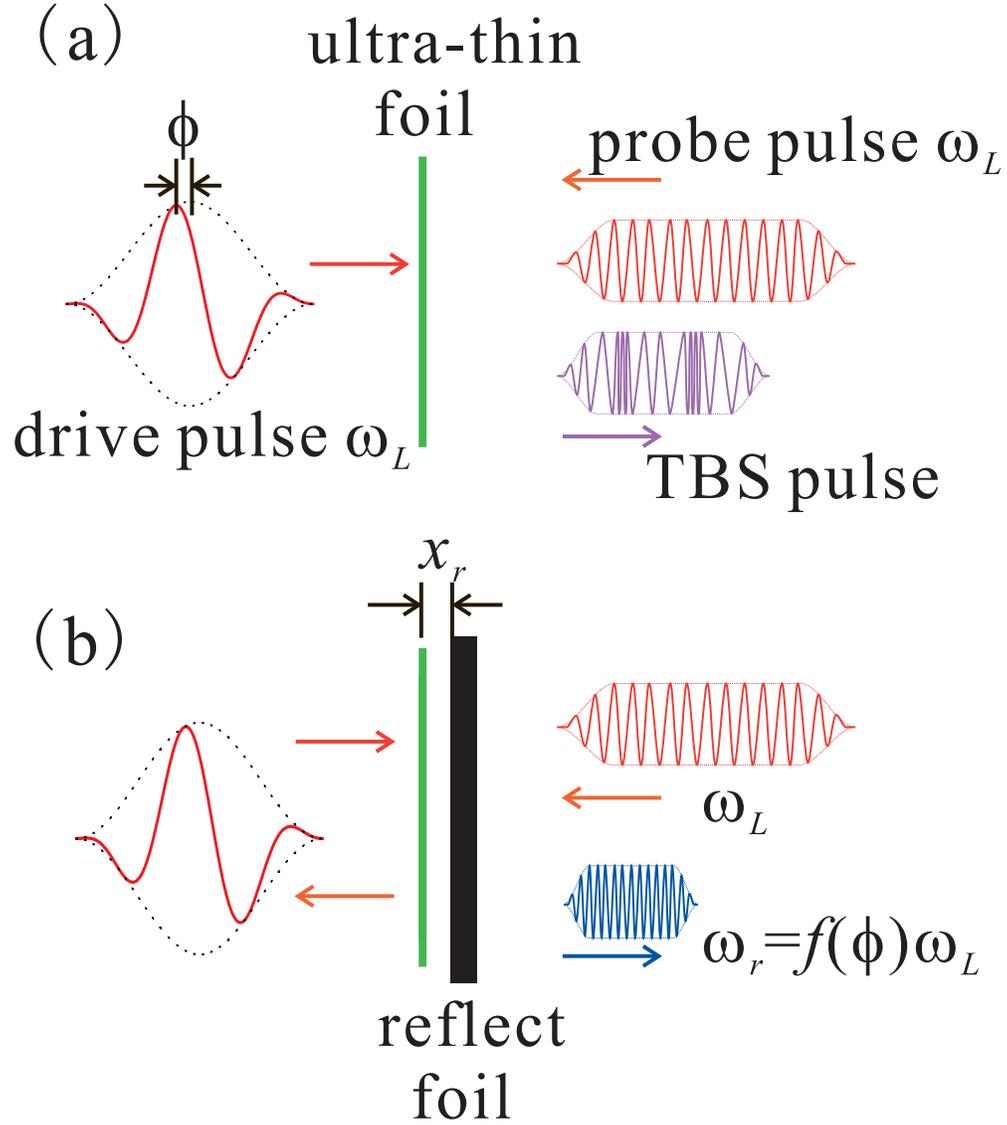}
\end{center}
\caption{\label{figure1}(Color online) (a) Schematic showing TBS of
a weak probe pulse by a flying mirror surfing on a relativistic
few-cycle pulse. The scattered pulse is strongly chirped due to the
acceleration of the electron layer. (b) Configuration of the CEP
measurement with the TBS light. The drive pulse accelerates the
flying mirror and is reflected by the reflect foil, without energy
consumed. After the relativistic flying mirror flies to the rear
side of the reflect foil, a counter propagating probe $\omega_0$
light is then mirrored and frequency upshifted to $ \omega  = f(\phi
)\omega _L $, which highly depends on the CEP of the drive pulse.}
\end{figure}

The corresponding configuration is sketched in
Fig.~\ref{figure1}(b). In this scheme, an intense few-cycle pulse
irradiates a target combined with an ultra-thin (nm) foil and a
thick and dense foil (the separation between these two foils is
$x_r$). The electrons of the ultra-thin foil driven by the intense
pulse play the role of a flying mirror. The thick foil behind will
reflect the drive pulse and let only the flying mirror pass through.
The flying mirror flies with a relativistic factor $\gamma _x = {1
\mathord{\left/
 {\vphantom {1 {\sqrt {1 - \beta _x^2 } }}} \right.
 \kern-\nulldelimiterspace} {\sqrt {1 - \beta _x^2 } }}
$, with $\beta _x  = {{v_x } \mathord{\left/
 {\vphantom {{v_x } c}} \right.
 \kern-\nulldelimiterspace} c}$ being the velocity of the plane
 flyer in the normal direction. A counter propagating probe light is
 then mirrored and frequency upshifted by the relativistic Doppler
 factor, which is
 ${{\left( {1 + \beta _x } \right)} \mathord{\left/
 {\vphantom {{\left( {1 + \beta _x } \right)} {\left( {1 - \beta _x } \right)}}} \right.
 \kern-\nulldelimiterspace} {\left( {1 - \beta _x } \right)}} \approx 4\gamma _x^2
$ for $\gamma _x  \gg 1$~\cite{Einstein,Meyer-ter-Vehn}. The
spectrum of the TBS light, achievable in experiments~\cite{Chang},
can be used to determine the CEP of the drive pulse.

For simple understanding, we start with only the ultra-thin foil
irradiated by an intense few-cycle laser pulse [see the left part of
Fig.~\ref{figure1}(a)].  The electrons gain high $\gamma$ values and
the heavy ions are left behind unmoved~\cite{Kulagin,Wen} when the
charge separation field is much smaller than the amplitude of the
laser field $E_{L0} $. The charge separation field depends on the
area charge density $\sigma _0=e n_0 d_0$, where $n_0$ and $d_0$ are
the plasma density and the foil thickness, respectively. In our
analytical model, all equations are presented in the nature unit.
The normalized quantities are obtained from their counterparts in SI
units marked with prime, i.e., time and length are normalized
according to $t=\omega_L t'$ and $l=k_L l'$, field $ E = {{eE'}
\mathord{\left/
 {\vphantom {{eE} {\left( {mc\omega _L } \right)}}} \right.
 \kern-\nulldelimiterspace} {\left( {mc\omega _L } \right)}} $,
 vector potential $ a = {{eA'} \mathord{\left/
 {\vphantom {{eA'} {\left( {mc} \right)}}} \right.
 \kern-\nulldelimiterspace} {\left( {mc} \right)}}
$, density $ n = {n' \mathord{\left/ {\vphantom {n' {n_c}}} \right.
 \kern-\nulldelimiterspace} {n_c}}$ and momentum $ p = {p' \mathord{\left/ {\vphantom
 {p'
{mc}}} \right.
 \kern-\nulldelimiterspace} {mc}}$, where $e$ and $m$ are the charge and the mass
 of the electron, $\omega_L$ and $k_L$ are the laser frequency
and the wave number, $c$ is the speed of light in vacuum and $ n_c
= {{\varepsilon _0 m\omega _L^2 } \mathord{\left/
 {\vphantom {{\varepsilon _0 m\omega _L^2 } {e^2 }}} \right.
 \kern-\nulldelimiterspace} {e^2 }}
$ is the electron critical density. We use a linearly polarized
($E_z=0$) pulse with a sine square envelop as the drive pulse
\begin{align}
E_y  = E_{y0} \sin ^2 \left( {{{\pi \tau } \mathord{\left/
 {\vphantom {{\pi \tau } T}} \right.
 \kern-\nulldelimiterspace} T}} \right) \cos \left( {\tau  + \phi } \right), \label{Ey}
\end{align}
with the propagating coordinate $\tau=t - x$ and the peak of
envelope $E_{y0}$. The dynamics are described by the
equations~\cite{Avetissian,Wen}
\begin{align}
\frac{{d\kappa }}{{d\tau }} = \left[ {\frac{{\sigma_0 }}{2\left( {1
+ p_y^2 } \right)}} \right], \label{dkappa}
\end{align}
\begin{align}
\frac{{dp_y }}{{d\tau }} =  - E_y  - \left[\frac{{\sigma_0
}}{2}\frac{{p_y }}{\kappa }\right]. \label{dpy}
\end{align}
Here the terms in the square brackets denote the self-fields of the
electron layer, and $E_y$ is the instantaneous laser field when the
electron layer surfing in the laser pulse. Analytically, $ \kappa =
\gamma  - p_x$ and $p_y$ can be obtained by integrating
Eqs.~(\ref{dkappa}) and~(\ref{dpy}) over the duration $[0, \tau]$.
With $\gamma ^2  = 1 + p_x^2  + p_y^2 $, one can find that the
energy of the flying mirror $\gamma$ is
\begin{align}
\gamma  = {{\left( {1 + p_y^2 } \right)} \mathord{\left/
 {\vphantom {{\left( {1 + p_y^2 } \right)} {2\kappa }}} \right.
 \kern-\nulldelimiterspace} {2\kappa }} + {\kappa  \mathord{\left/
 {\vphantom {\kappa  2}} \right.
 \kern-\nulldelimiterspace} 2}.
 \label{gamma}
\end{align}
Although the self-radiation and the charge separation field are
taken into account in this model, the dominating force is still from
the drive pulse. When the charge surface density $ \sigma _0 $ is
considerably small compared with the laser field $E_{L0}$, the
analytical model will regress to single electron model $\kappa \to
1$ and the energy gain of electron layer is proportional to the
square of instantaneous vector potential of the drive pulse
\begin{align}
\begin{array}{c}
 \left( {\Delta \gamma } \right)^ +   = {{p_y^2 } \mathord{\left/
 {\vphantom {{p_y^2 } 2}} \right.
 \kern-\nulldelimiterspace} 2} \approx {{\left[ {\int_0^\tau  {E_y \left( {\tau ,\phi } \right)d\tau } } \right]^2 } \mathord{\left/
 {\vphantom {{\left[ {\int_0^\tau  {E_y \left( {\tau ,\phi } \right)d\tau } } \right]^2 } 2}} \right.
 \kern-\nulldelimiterspace} 2} \\
  = {{a_y^2 \left( {\tau ,\phi } \right)} \mathord{\left/
 {\vphantom {{a_y^2 \left( {\tau ,\phi } \right)} 2}} \right.
 \kern-\nulldelimiterspace} 2} \propto \cos \left[ {2\phi  + g\left( \tau  \right)} \right], \\
 \end{array}
\label{dgamma}
\end{align}
where $g\left( \tau  \right)$ is a function of $\tau$.
Equation~(\ref{dgamma}) shows the energy gain of the flying mirror
$\left( {\Delta \gamma } \right)^ +$ varies periodically with the
CEP of the drive pulse. In other words, a shift of $\pi$ in the CEP
would induce the same results.  It shows the energy of the flying
mirror is mainly dependent on the temporally varied vector potential
$a_{y}$, and carries the detailed information of the drive pulse.
Through a proper process to obtain the energy of the flying mirror
(TBS shown later), we can extract the CEP of the laser pulse.

\begin{figure}
\begin{center}
\includegraphics[width=1.0\columnwidth]{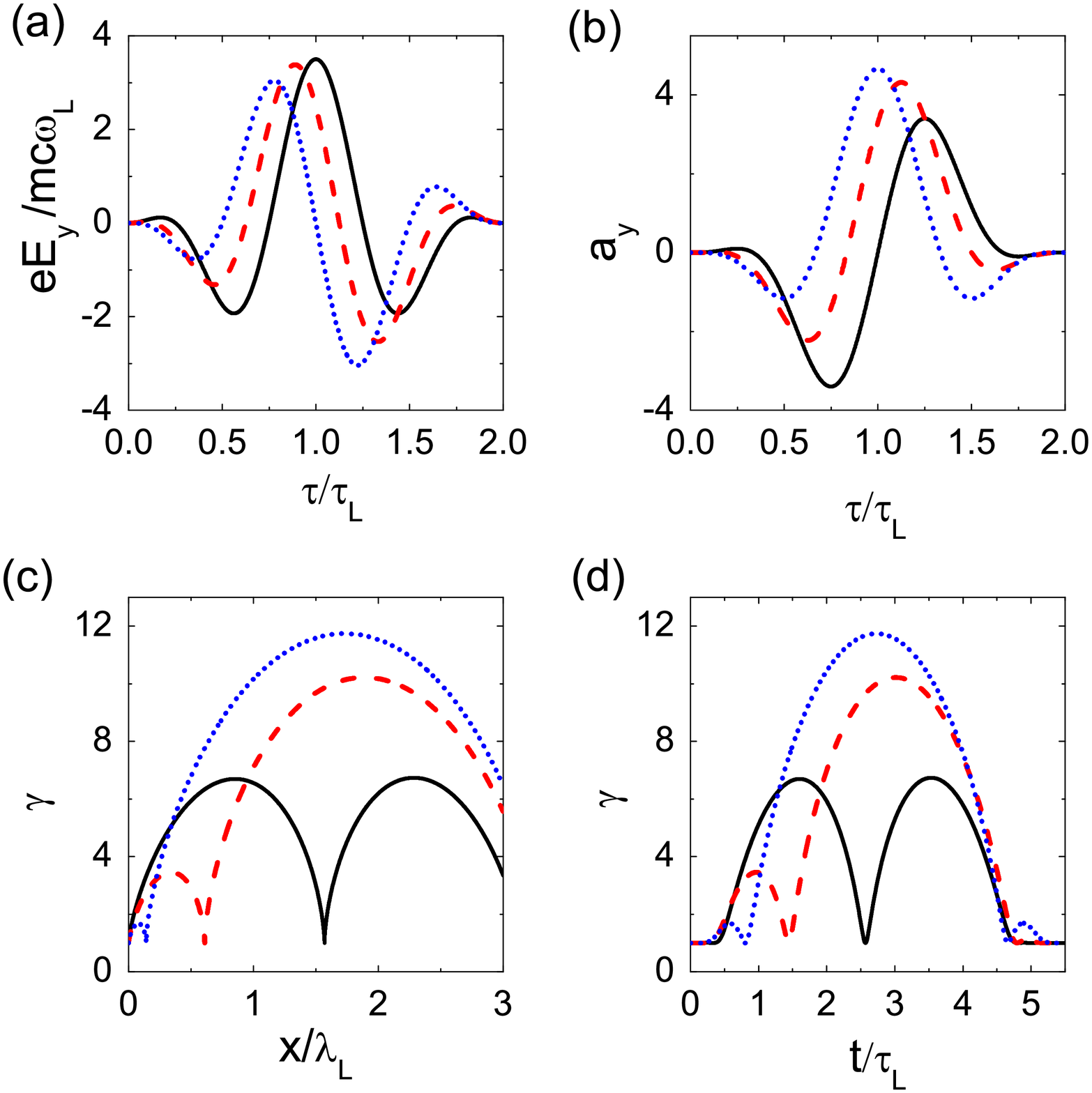}
\end{center}
\caption{\label{figure2}(Color online) (a) The electric field, (b)
vector potential, (c) spatial and (d) temporal evolution of
normalized electron layer energy in the few cycle laser field with
$\phi=0$ [solid (black) curve], $\pi/4$ [dashed (red) curve] and
$\pi/2$ [dotted (blue) curve] (dotted curve) obtained from
analytical model. }
\end{figure}

We consider a relativistically intense laser field with a peak
amplitude of $E_{y0}=3.5$, corresponding to an intensity of $ I =
2.6 \times 10^{19} {{\rm{W}} \mathord{\left/
 {\vphantom {{\rm{W}} {{\rm{cm}}^{\rm{2}} }}} \right.
 \kern-\nulldelimiterspace} {{\rm{cm}}^{\rm{2}} }} $ for $\lambda_L  =
 800{\rm{nm}}$, with pulse duration $T = 2 \tau_L$, where $\lambda_L$
 and $\tau_L$ are laser wavelength and period. The numerical results from our analytical model are plotted in
Fig.~\ref{figure2}.
 Figures~\ref{figure2}(a) and (b) show that the temporal
 variation of the electric field and the vector potential depend on
the CEP. The energy of the flying mirror during the few-cycle laser
field with different CE phases are shown in Figs.~\ref{figure2}(c)
and (d). Figure~\ref{figure2}(c) exhibits how the energy of the
flying mirror evolves along the laser propagation $x$. It is shown
that the peak value of $\gamma$ depends on the CEP of the
relativistically intense laser. The maximal energy of flying mirror
can be almost doubled by choosing the CEP properly, e.g.,
$\gamma_{max}=6.7$ when $\phi=0$ (solid curve), while
$\gamma_{max}=11.8$ when $ \phi = {\pi \mathord{\left/
 {\vphantom {\pi  2}} \right.
 \kern-\nulldelimiterspace} 2}$ (dotted curve). If we detect the energy at a fixed position $x_0$, it varies
periodically with the CEP of the drive laser pulse. A similar trend
appears in the dependence of the electron layer energy on time $t$,
as shown in Fig.~\ref{figure2}(d).

\begin{figure}
\begin{center}
\includegraphics[width=1.0\columnwidth]{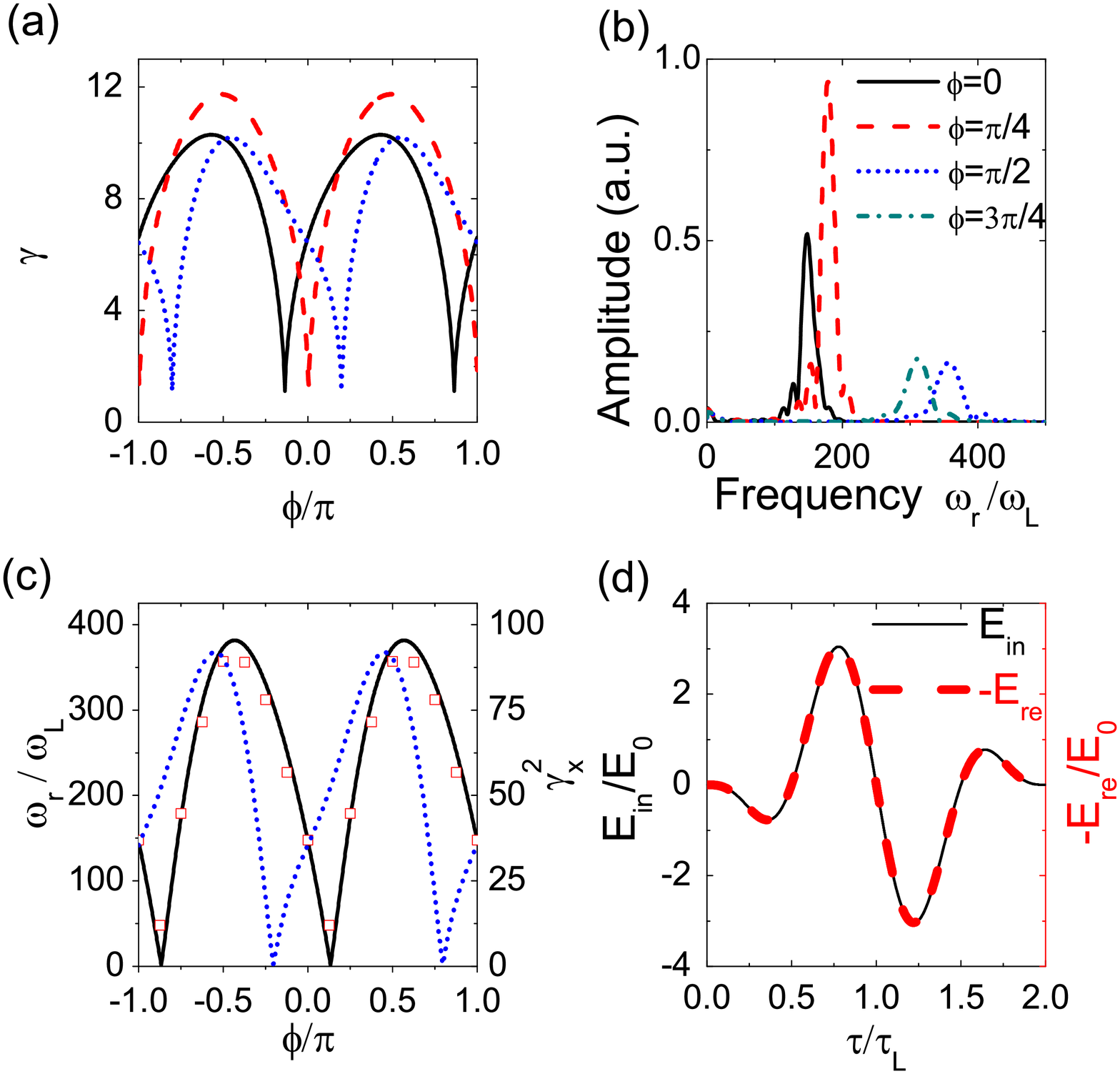}
\end{center}
\caption{\label{figure3}(Color online) (a) The normalized energy of
electron layer $\gamma$ as functions of the
 initial CEP of the drive ultrashort pulse, when the electron
 layer reaches the location of the reflect foil $x_r$. The solid (black),
 dashed (red) and dotted (blue) curves correspond to $x_r=\lambda,~1.6\lambda$ and $2.5\lambda$, respectively.
(b) The spectra of Thomson backscattering light with the CE phases
of drive pulses $ \phi =0,~ \pi/4,~ \pi/2$ and $ 3\pi/4$ for
$x_r=\lambda$. (c) When $x_r=\lambda$ the central frequency of the
TBS light as a function of the CEP obtained from the model [(black)
solid curve], with the matched simulation results [(red) open
cubes). The (blue) dotted curve represent the dependence of the
detected central frequency on the CEP of the drive pulse when
$x_r=2.5\lambda$. The electric fields of the incident and reflected
drive pulse are compared in (d).}
\end{figure}

Bright VUV- or X-ray source can be obtained by TBS from the
relativistic flying mirror. It has been demonstrated that the TBS
light from a flying mirror is chirped and has a broad
spectrum~\cite{Meyer-ter-Vehn,Wu}, sketched in
Fig.~\ref{figure1}(a), which makes it difficult to find a central
frequency in the spectrum of the TBS light. However, this can be
overcome by setting a thick foil as a reflector behind the
ultra-thin foil with a distance $x_r$~[see Fig.~\ref{figure1}(b)],
which is a practical way to generate a uniform flying
mirror~\cite{Wu,Wang}. After the flying mirror emerges from the
reflect foil and divorces from the drive laser, its energy $\gamma$
becomes a constant and depends on the CEP of the pulse [see
Fig.~\ref{figure3}(a)]. The solid (blue), dashed (green) and dotted
(red) curves correspond to $x_r=\lambda$,
 $x_r=1.6\lambda$ and $x_r=2.5\lambda$, respectively. The same with the analytical
prediction from Eq.~(\ref{dgamma}), $\gamma$ is a periodic function
of the CEP $\phi$ with a period of $\pi$.

When a probe light irradiates this flying mirror, the frequency of
the probe pulse is upshifted by a fixed factor $ {\omega_r
\mathord{\left/
 {\vphantom {\omega  {\omega _L }}} \right.
 \kern-\nulldelimiterspace} {\omega _L }} = 4\gamma _x^2 $, and
 an isolated pulse with a narrow spectrum is generated. We verify the
 results by 1D-PIC simulations~\cite{Sheng}. The foil parameters in the
 simulations are the same as those in Ref.~\cite{Wu}, i.e., density $n_0/n_c=1$ and
 thickness $d_0/\lambda_L=0.001$ for the ultra-thin foil, and $n_1/n_c=400$
and $d_1/\lambda_L=0.1$ for the reflect foil. The simulation results
of the TBS light spectra with different CE phases
  of the drive pulse are shown in Fig.~\ref{figure3}(b) when
$x_r=\lambda$. We choose $ \phi =0$, $ \phi  = {\pi  \mathord{\left/
 {\vphantom {\pi  4}} \right.
 \kern-\nulldelimiterspace} 4}$, $
\phi  = {\pi  \mathord{\left/
 {\vphantom {\pi  2}} \right.
 \kern-\nulldelimiterspace} 2}$ and $
\phi  = {{3\pi } \mathord{\left/
 {\vphantom {{3\pi } 4}} \right.
 \kern-\nulldelimiterspace} 4}$ for examples. As a result of
the dependence of the flying mirror energy on the CEP, the central
frequency of the spectrum is sensitive to the CEP, e.g., varying
from $\sim148\omega_L$ at $\phi=0$ to $\sim357\omega_L$ at
$\phi=\pi/2$. The central frequency of the TBS light as a periodic
function of the CEP predicted by the analytic model is shown in
Fig.~\ref{figure3}(c).

It should be noticed that the reflected intense light would interact
with the flying mirror and cause an energy loss $ \left( {\Delta
\gamma } \right)^ - \approx {1 \mathord{\left/
 {\vphantom {1 2}} \right.
 \kern-\nulldelimiterspace} 2} $, although the interaction is extremely short as
compared with the acceleration during the co-moving process with the
drive pulse for a long time~\cite{Meyer-ter-Vehn}. After the flying
mirror goes through the reflector, its transverse momentum is
canceled $p_y=0$, and a very uniform relativistic flying mirror
[with the energy $\gamma_x=\gamma-\left( {\Delta \gamma } \right)^
-$] is obtained, while the relative energy loss via Coulomb collsion
is found to be negligible~\cite{Wu}. Taking the CEP of the drive
laser into consideration, $\gamma_x$ depends periodically on the CEP
$\phi$, illustrated by the right axis in Fig.~\ref{figure3}(c).
Obeying the relation $ {\omega_r \mathord{\left/
 {\vphantom {\omega  {\omega _L }}} \right.
 \kern-\nulldelimiterspace} {\omega _L }} = 4\gamma _x^2 $, the central
frequency also exhibits periodicity on the CEP.
The analytical predictions agree well with the simulation
results.

Due to the period of $\pi$ we will get two possible phases with one
measurement, e.g. $\omega_r = 200 \omega_L$ while $\phi_1= 0.27\pi$
and $\phi_2=0.94\pi$ ($x_r=\lambda$). By introducing a second
measurement, this restriction can be removed and the CEP can be
determined in the range of $\pi$. The simulations show when the
drive pulse is highly reflected with a limited energy loss [shown in
Fig.~\ref{figure3}(d)], the drive pulse can be transmitted to
another double-foil target again with a different distance $x_r$.
For example, if the first measurement gives $\omega_r^{1st} = 200
\omega_L$, for the second measurement with $x_r=2.5\lambda$, the CEP
is determined to be $0.27\pi$ when second measurement gives
$\omega_r^{2nd} = 297 \omega_L$, or $0.94\pi$ when $\omega_r^{2nd} =
115 \omega_L$. Moreover, the central frequency is very sensitive to
the CEP. For example, around the point $\left( {\left. {0.27\pi }
\right|_\phi  ,\left. ~ {200\omega _L } \right|_{\omega _r } }
\right)$, a difference of $1\omega_L$ in the detectable central
frequency introduces a phase shift of $ 8 \times 10^{ - 4} \pi $ in
the CEP.

In summary, the evolution of a flying mirror driven by a
relativistic, few-cycle pulse ($ I = 2.6 \times 10^{19} {{\rm{W}}
\mathord{\left/
 {\vphantom {{\rm{W}} {{\rm{cm}}^{\rm{2}} }}} \right.
 \kern-\nulldelimiterspace} {{\rm{cm}}^{\rm{2}} }} $ and $T \approx 5.3 \rm{~fs}$ at $\lambda
= 0.8 \rm{~\mu m}$) from an ultra-thin foil is investigated
theoretically. With the help of a reflect foil, a TBS light pulse
with narrow spectrum is obtained when a probe light is reflected
from the flying mirror. The central frequency of the TBS light is a
periodic function of the CEP, with the period of $\pi$. The
detection of the central frequency of the TBS light makes it
possible to determine the CEP of a relativistic few-cycle pulse. We
introduce a double-measurement process to determine the CEP in the
range of $\pi$. In principle, this method is also feasible for a
weaker or longer pulses with relativistic intensity ($I > 10^{18}
{{\rm{W}} \mathord{\left/
 {\vphantom {{\rm{W}} {{\rm{cm}}^{\rm{2}} }}} \right. \kern-\nulldelimiterspace} {{\rm{cm}}^{\rm{2}} }}$), while
 even thinner foil is needed to generate a uniform flying mirror.

The authors are grateful to Dr. A. Di Piazza for useful discursion.
This work was supported by National Nature Science Foundation of
China (Grant Nos. 10935002, 11025523, 61008016 and 10905003) and
National Basic Research Program of China (Grant No. 2011CB808104).
M.Wen and L.L.Jin acknowledges the support from the Scientific
Research Program Funded by Shaanxi Provincial Education Department
(Program Nos. 11JK0538 and 11JK0529).

\bibliography{apssamp}

\end{document}